\documentclass[aps,pra,superscriptaddress,12pt,tightenlines,nofootinbib]{revtex4}
\usepackage{amsmath,amsthm,graphicx,amssymb,verbatim,listings}
\usepackage{wasysym}
\usepackage[T1]{fontenc}     % needed for the guillemets
\usepackage{lmodern}         % needed to fix suboptimal font rendering
\usepackage[ pdftex, plainpages = false, pdfpagelabels,
                 pdfpagelayout = useoutlines,
                 bookmarks,
                 bookmarksopen = true,
                 bookmarksnumbered = true,
                 breaklinks = true,
                 linktocpage=all,
                 pagebackref=false,
                 colorlinks = true,
                 linkcolor = BrickRed,
                 urlcolor  = blue,
                 citecolor = BrickRed,
                 anchorcolor = green,
                 hyperindex = true,
                 hyperfigures
                 ]{hyperref}
\usepackage[usenames, dvipsnames]{xcolor}
\usepackage{xifthen} % provides \isempty test

\newcommand{\tr}{\hbox{tr}}

\newcommand{\arxiv}[2][]{\ifthenelse{\isempty{#1}}{\href{http://arxiv.org/abs/#2}{{\tt arXiv:\allowbreak{}#2}}} {\href{http://arxiv.org/abs/#2}{{\tt arXiv:\allowbreak{}#2 [#1]}}}}

\newcommand{\booktitle}{\textsl}
\newcommand{\hrefdoi}[2]{\href{https://doi.org/#1}{#2}}

\newtheorem{earman}{Earmanism}
\newcommand{\bje}{\begin{earman}} %\protect$\!\!${\em\bf :}$\;\;$}
\newcommand{\eje}{\end{earman}}

\begin{document}

\title{QBians Do Not Exist}

\author{Christopher A.\ Fuchs}
\author{Blake C.\ Stacey}
\affiliation{\href{http://www.physics.umb.edu/Research/QBism/}{Physics Department}, University of Massachusetts Boston}

\date{\today\bigskip}

\bigskip

\begin{abstract}
We remark on John Earman's paper ``Quantum Bayesianism Assessed'' [\booktitle{The Monist} \textbf{102} (2019), 403--423], illustrating with a number of examples that the quantum ``interpretation'' Earman critiques and the interpretation known as QBism have almost nothing to do with each other.
\end{abstract}

\maketitle

\section{Introduction}

\begin{flushright}
\small``PROBABILITY DOES NOT EXIST''\\
--- Bruno de Finetti, 1990 \cite[p.\ x]{deFinetti1990}
\medskip\\
\small ``Bruno de Finetti \ldots\ proclaimed `THERE ARE NO PROBABILITIES' (de Finetti 1990, x)''\\
--- John Earman~\cite{Earman:2019}
\end{flushright}

Philosophers pride themselves on being careful readers and analyzers.  In a recent article in {\it The Monist}, ``Quantum
Bayesianism Assessed''~\cite{Earman:2019}, John Earman analyzes a position he ascribes to a group of people he calls ``QBians''---a term meant to suggest followers or contributors to the quantum interpretative project initially known as Quantum Bayesianism~\cite{Caves02a}, but now after a significant number of modifications, simply called QBism~\cite{Fuchs:2010,FMS,FuchsStacey2018}.\footnote{For an introduction to the origins of QBism, see BCS's ``Ideas Abandoned en Route to QBism,'' \cite{Stacey:2019b}.}  In another article~\cite{Earman18}, Earman even refers to ``Quantum Bayesians (QBians as they style themselves).''  As a point of fact, no QBist (the moniker {\it that is used\/} by followers of QBism\footnote{The term QBist appears to be traceable to this 2009 paper~\cite{Fuchs09}, which was later refined into the 2013 \booktitle{Reviews of Modern Physics} article~\cite{Fuchs13a}.}) has ever identified with this alternative term, and no previous commentator on QBism has ever employed it. QBians do not exist; it is a coinage of Earman alone.

In a way this is fortunate, as the portrayal Earman gives has so little to do with the actual language, goals, metaphysics, and mathematical technicalities of QBism as to be unrecognizable.  Even the current Wikipedia article on QBism is truer to its subject~\cite{Wiki}:
\begin{quotation}
\noindent According to QBism, quantum theory is a tool which an agent may use to help manage his or her expectations, more like probability theory than a conventional physical theory. Quantum theory, QBism claims, is fundamentally a guide for decision making which has been shaped by some aspects of physical reality. Chief among the tenets of QBism are the following:
\begin{enumerate}
\item
    All probabilities, including those equal to zero or one, are valuations that an agent ascribes to his or her degrees of belief in possible [measurement] outcomes. [From later in the article:  QBism considers a measurement to be any action that an agent takes to elicit a response from the world and the outcome of that measurement to be the experience the world's response induces back on that agent.]  As they define and update probabilities, quantum states (density operators), channels (completely positive trace-preserving maps), and measurements (positive operator-valued measures) are also the personal judgements of an agent.
\item
    The Born rule is normative, not descriptive. It is a relation to which an agent should strive to adhere in his or her probability and quantum state assignments.
\item
    Quantum measurement outcomes are personal experiences for the agent gambling on them. Different agents may confer and agree upon the consequences of a measurement, but the outcome is the experience each of them individually has.
\item
    A measurement apparatus is conceptually an extension of the agent. It should be considered analogous to a sense organ or prosthetic limb---simultaneously a tool and a part of the individual.
\end{enumerate}
\end{quotation}
But this also means that Earman's paper is an opportunity lost, as it becomes essentially irrelevant to any meaningful discussion on the difficulties of QBism.  (Any QBist would agree that QBism is an unfinished project in need of thoughtful input.)

To take some initial examples, consider the following quotes from Earman's paper:
\bje
The elements of the projection lattice $\mathcal{P}(\mathfrak{B}(\mathcal{H}))$ are referred to as quantum propositions (also yes-no questions, or quantum events). Quantum probability theory may be thought of as the study of quantum probability functions {\rm Pr} on $\mathcal{P}(\mathfrak{B}(\mathcal{H}))$.
\eje
\bje
For Bayesians, classical or quantum, probability is degree of belief,
the objects of which are propositional entities. In contrast to the
projection lattice $\mathcal{P}(\mathfrak{B}(\mathcal{H}))$, the
effect algebra does not have a natural propositional structure.
\eje
\bje
Suppose that a QBian agent learns that $F \in
\mathcal{P}(\mathfrak{B}(\mathcal{H}))$ is true.
\eje
None of this coincides with the way QBists think of probability's usage within quantum theory.  All one need do is look through any of the QBist and proto-QBist papers Earman cites in his article, Refs.~\cite{Caves02a,Fuchs:2010,FMS,Caves02c,Caves02b,Caves07,Fuchs:2002,Fuchs04b,Fuchs13a,Fuchs04a,Mermin12a,Mermin14c,Schack01a,vonBaeyer16a}, or even the critical assessment by Timpson~\cite{Timpson08a} which he also cites. One will search in vain to find any use made of a non-Boolean ``lattice of propositions'' or any hint that QBists subscribe to a notion that upon measurement one ``learns the truth value'' of a proposition.

Indeed already as of 2002, one of us (CAF) was writing consistently \cite{Fuchs:2002} of an opposing conception of measurement:
\begin{quotation}
\noindent It is a theory not about observables, not about beables, but about
``dingables.'' We tap a bell with
our gentle touch and listen for its beautiful ring.

So what are the ways we can intervene on the world?  What are the
ways we can push it and wait for its unpredictable reaction?  The
usual textbook story is that those things that are measurable
correspond to Hermitian operators.  Or perhaps to say it in more
modern language, to each observable there corresponds a set of
orthogonal projection operators $\{\Pi_i\}$ over a complex Hilbert
space ${\cal H}_{\scriptscriptstyle\rm D}$ that form a complete
resolution of the identity \ldots

Nonetheless, one should ask: Does this theorem [Gleason's theorem]
really give {\it the physicist\/} a clearer vision of where the
probability rule comes from? \ldots

The place to start is to drop the fixation that the basic set of
observables in quantum mechanics are complete sets of orthogonal
projectors.  In quantum information theory it has been found to be
extremely convenient to expand the notion of measurement to also
include general positive operator-valued measures (POVMs) \ldots
\end{quotation}
And if one digs deeper into the literature, one will easily find discussions like this~\cite{Samizdat2,Fuchs20}:
\begin{quotation}
\noindent It is not that I reject a well-structured event space because it {\it assumes\/} its elements would correspond to intrinsic properties of quantum systems, but rather this is the {\it result\/} of a thoroughgoing subjective interpretation of probabilities within the quantum context.  What cannot be forgotten is that quantum-measurement outcomes, by the usual rules, {\it determine\/} posterior quantum states.  And those posterior quantum states in turn determine further probabilities.

[I]f one takes the timid, partial move that [Pitowsky] and [Bub], say, advocate---i.e., simply substituting one or another non-Boolean algebra for the space of events, and leaving the rest of Bayesian probability theory seemingly intact---then one ultimately ends up re-objectifying what had been initially supposed to be subjective probabilities.  That is:  When I look at the click, and note that it is value $i$, and value $i$ is rigidly---or I should say, {\it factually}---associated with the projector $\Pi_i$ in some non-Boolean algebra, then I have no choice (through L\"uders rule) but to assign the posterior quantum state $\Pi_i$ to the system.  This means the new quantum state $\Pi_i$ will be as factual as the click.  And any new probabilities (for the outcomes of further measurements) determined from this new quantum state $\Pi_i$ will also be factual.

So, the starting point of the reasoning is to {\it assume\/} that there is a category distinction between probabilities and facts (this is the subjectivist move of de Finetti and Ramsey).  Adding the ingredient of the usual rules of quantum mechanics, one derives a dilemma:  If there is a rigid, factual connection between the clicks $i$ and elements $\Pi_i$ of an algebraic structure, then probabilities are factual after all.  Holding tight to my assumption of a category distinction between facts and probabilities, I end up rejecting the idea that there is a unique, factual mapping between $i$ and $\Pi_i$. \ldots

Though [QBists] banish the algebraic structure of Hilbert space from having anything to do with a fundamental event space (and in this way their quantum Bayesianism differs from the cluster of ideas Pitowsky and Bub are playing with), they do not banish the algebraic structure from playing any role whatsoever in quantum mechanics.  It is just that the algebraic structure rears its head at the conceptual level of coherence rather than in a fundamental event space.  It is not that potential events are objectively tied together in an algebraic way, but that our gambling commitments (normatively) {\it should be}.
\end{quotation}
This, for instance, is a good bit of the reason that the technical side of the QBist program~\cite{Fuchs13a,Fuchs11,Appleby11b,Appleby11c,Appleby13,Appleby15,Appleby:2017} focusses on trying to recover the algebraic structure of quantum theory from a primitive (nonalgebraic) statement of the Born rule, rather than the other way around, as the Gleason approach does.  Where Earman writes in his paper,
\bje
Some QBians struggle mightily over the status of the `Born rule.' \ldots\ But there is no mystery here \ldots\ nor is extra normative guidance required. By Gleason's theorem a countably additive (respectively, completely additive) measure {\rm Pr} on $\mathcal{P}(\mathfrak{B}(\mathcal{H}))$ with $\mathcal{H}$ separable \ldots\ whether or not it is given a personalist interpretation, corresponds to a normal state \ldots.
\eje
\noindent it can only mean that he has not absorbed or internalized what the program is about.  

Despite Earman's assertions, QBists do not ``struggle mightily'' over the status of the Born rule:  It is the centerpiece of their technical research program. QBists believe that, when expressed more primitively as a relation between probabilities (rather than as a hybrid of operators and probabilities), the Born rule exhibits the most incisive statement yet on how quantum probability departs from classical theory~\cite{DeBrota20b}. Thus it is the most promising place to look for the deepest lesson quantum theory is yet to reveal about physical reality. This outlook contrasts starkly with other interpretations of quantum mechanics, which typically want to take the Born rule as an afterthought to be derived or an embarrassment to be elided.  How many times do the QBists have to say in print that QBism is seeking an inversion of what Gleason and Mackey were aiming for?

In summary, Earman's QBian is a straw man---a creation of his own, not the literature he cites.  Nothing says this more tellingly than,
\bje
QBians occasionally make use of this operational approach [i.e., the framework of POVMs], but it is unclear how it lends itself to something that deserves to be called quantum Bayesianism.
\eje
``Occasionally''?  It was the foremost notion in nearly every paper Earman cited!  ``[I]t is unclear how it lends itself to something that deserves to be called quantum Bayesianism''?  The only conclusion one can draw is that the author simply did not do his homework before declaring in print what he thinks QBism {\it ought\/} to be.

So it is that the term QBian is a fortunate choice to name this arbitrary creation.  On the other hand, many a young philosopher will look up to the writings of the distinguished professors in their field as entry points for their own thinking.  A philosopher is certainly more likely to first learn of QBism through another philosopher than to have stumbled across the original physics literature.  Therein lies the danger of an article like Earman's.\footnote{One of us (CAF) has witnessed this firsthand, in correspondence from PhD students at the University of Oxford and the London School of Economics, querying him about this or that ``QBian'' stance, not realizing the idiosyncracy of the term and how disconnected Earman's commentary on QBism is from the actual view.}

What can be done about this?  Does Earman's paper deserve a careful rebuttal?  We wouldn't even know how to start, so disconnected the paper is from actual QBism.  It does seem worthwhile, however, to make enough limited remarks on some of Earman's passages to sow a seed of doubt and suggest that QBism is something else than he portrays it to be.  The remainder of our paper is devoted to such an exercise.

\section{A Historical Note}

\bje
But in recent years the major push for a personalist reading of
quantum probabilities has come not from philosophers but from
physicists who march under the banner of quantum Bayesianism (or QBism
for short). In the vanguard are Carleton Caves, Christopher Fuchs,
R\"udiger Schack, and David Mermin.
\eje

The view espoused by Caves, Fuchs and Schack~\cite{Caves02a} in the early 2000s, while it can be called a kind of ``Quantum Bayesianism'', is
not QBism~\cite{Fuchs:2010}. The latter grew out of the former and rejects many
statements made under the earlier banner~\cite{Stacey:2019b}. Carlton
Caves in fact does not consider himself a QBist~\cite{Stacey:2019b} (and \emph{Carlton} is not spelled with an E). To
define QBism, Earman cites several papers from 2001 through 2014, not
distinguishing the very different philosophies they represent. The
cited papers do not form a coherent whole. The earlier ones say much
that the later ones contradict, and the later publications explicitly
disavow the earlier. This is a consequence of progress (the sort of thing physicists strive for). We, of
course, believe that it is progress toward physical understanding, but
we hope that even non-QBists can see it as advancing toward internal
self-consistency.

Earman refers to a pair of completely different
papers~\cite{Fuchs:2002, Fuchs:2010} as ``Versions of Fuchs's QBian
manifesto''. Of course, neither is ``QBian''; moreover, the article
from 2002 is still a significant phase transition away from QBism proper, not yet
embracing the \emph{two levels of personalism}~\cite[Introduction]{Samizdat2} that Fuchs and Schack
had developed by 2010 in order to resolve the paradox of Wigner's
Friend~\cite{Stacey:2019b,Fuchs11b,Fuchs14,DeBrota:2020}.

\begin{comment}
In a pre-publication version of Earman's article, there is a passage
in which he devises an interpretation that is not QBism, which he then
proceeds to criticize instead of critiquing QBism.
\bje
Quantum probability is here construed as the study of probability
functions $pr$ on the lattice of quantum propositions
$\mathcal{P}(\mathfrak{N})$ about the system described by
$\mathfrak{N}$. [\ldots] For present purposes I propose that quantum
Bayesianism is built on the form of Bayesian personalism for quantum
propositions sketched above---probability functions on
$\mathcal{P}(\mathfrak{N})$ construed as credence functions of
rational agents who update by L\"uders conditionalization.
\eje
The published version of the paper omits the admission that it is
arguing against an invented position, while not actually changing its
arguments. The emphases of this passage ---
lattices of ``quantum propositions'', the L\"uders rule --- carry over
to the published version.

\end{comment}

\section{On Lattices and Logic}
The position that Earman invents resembles a probabilistic gloss on
quantum logic. Indeed,
it directly contradicts well-established features of QBism in multiple
evident ways. For example, see Earmanisms 1--3 above.  In QBism,
performing a measurement is not a matter of learning that a
proposition is true. Measurements do not reveal pre-existing
properties of the measured object; nor do they merely instill
properties within the object. Instead, a measurement outcome is an
experience~\cite{Fuchs15b,Fuchs2017}, the \emph{agent's side} of a
joint agent-object event.\footnote{Predecessors to this view include
  Wheeler's ``elementary acts of
  observer-participancy''~\cite{Wheeler:1982,Fuchs16}. We have also noticed a
  family resemblance to aspects of Whitehead's process
  philosophy~\cite{Whitehead:1979} and Barad's
  ``intra-actions''~\cite{Barad:2007}. However, QBism aims to develop
  these kinds of imagery more exactly and quantitatively than has
  ever been done before.}

Because QBism does not treat measurement outcomes as propositions that
are true- or false-valued, it has no fundamental role for a lattice of
projections.  Instead, a QBist assigns POVM elements (or ``effects'')
to potential future experiences, and then associates a consistent set
of personal probabilities to those POVM elements.  Both the
probabilities and the choice of POVM elements are personal judgments,
expressions from the agent's own mesh of beliefs cashed out as
gambling commitments. Quantum measurements are not restricted to the
narrow notion of orthonormal bases, and even when POVM elements are
rank-1 projection operators, they do not stand for ``quantum
propositions''. Consequently, Earman's language is inapplicable, and
his mathematical requirements are unmotivated. The effect algebra is a
natural and physically meaningful entity, as decades of daily work in
quantum information has proved; if it lacks a propositional structure,
so much the worse for the notion of propositional structure.

In QBism, noncommuting operators and non-Boolean lattices are
secondary or tertiary consequences of more fundamental physical
principles~\cite{Fuchs20}. Earman's failure to recognize this has
consequences. For example, he writes,
\bje
But the failure of Gleason's theorem [in dimension 2] should be a
cause for alarm on the bottom-up personalist reading of probabilities.
\eje
The analogue of Gleason's theorem holds in dimension 2 if the set of
quantum measurements is the set of POVMs, rather than that of von
Neumann measurements~\cite{Busch:2003}. Without the antiquated
quantum-logic drive to find the latter fundamental, there is no reason
to regard the case of dimension 2 as exceptional or problematic. This
observation, made independently by Busch and by Caves et al., predates
QBism proper by several years~\cite{Caves:2004}.

Earman discusses the L\"uders conditionalization
rule~\cite{Barnum:2000, Busch:2009} at some length. A QBist agent is
not constrained to use the L\"uders rule. Instead, a QBist regards the
L\"uders rule as a sensible way that an agent \emph{might expect to
  update her beliefs} under some conditions, not as a defining feature
of rationality or anything like that~\cite{Fuchs:2012, DeBrota:2019}.
The standard formalism for representing quantum-state change is that
of Kraus operators, which allow an agent to associate a different
completely positive trace-preserving map to each outcome of a
measurement~\cite{Wilde:2017}. To a QBist, the choice of assigning a
particular POVM $\{E_i\}$ to represent a measurement is a personal
judgment, and so is the choice of Kraus decomposition
\begin{equation}
  E_i = \sum_j A^\dag_{ij} A_{ij}
\end{equation}
for each POVM element. Consequently, all of the ingredients in the
state-change rule
\begin{equation}
  \rho \to \rho' = \frac{1}{\tr \rho E_i} \sum_j  A_{ij} \rho A_{ij}^\dag
\end{equation}
have the same conceptual status. Whether or not an agent chooses to
have this reduce to the L\"uders form
\begin{equation}
  \rho \to \rho' = \frac{\Pi_i \rho \Pi_i}{\tr \rho \Pi_i},
\end{equation}
where $\{\Pi_i\}$ is a set of projectors onto an
orthonormal basis for $\mathcal{H}$, depends on that agent's personal
mesh of beliefs. First, the agent would have to choose to associate a
particular set of potential future experiences with the specific
projectors $\{\Pi_i\}$, and then, the agent would have to decide that
applying the L\"uders rule is consistent with their belief
mesh. Neither step is obligatory. For example, having chosen a POVM
$\{E_i\}$ as the mathematical model of her action, an agent may
invoke a generalized version of the L\"uders rule,
\begin{equation}
  \rho \to \rho' = \frac{1}{\tr \rho E_i}
  \sqrt{E_i} \rho \sqrt{E_i},
\end{equation}
or she may invoke an ``entanglement-breaking''~\cite{Ruskai:2003,
  DeBrota:2020b, Pandey:2020} update rule
\begin{equation}
  \rho \to \rho' = \frac{E_i}{\tr E_i}.
\end{equation}
These will in general differ.

QBism subscribes to a school of personalist probability \cite{Hacking67,vanFraassen84} in which an
agent need not always update by the Bayes conditioning rule, or even
expect that she will~\cite{Fuchs:2012}. Consequently, the question
that Earman raises of when L\"uders conditionalization can be made to
resemble the Bayes rule is, on a basic conceptual level, doubly
irrelevant.

Earman regards the subject of quantum-state preparation as vitally
significant. He writes, for example,

\bje
From the objectivist perspective Prop.\ 3 can be viewed as a special
case of Lewis's PP [Principal Principle]: when an agent learns that
$S_\varphi$ is true she learns that the objective chances are given by
the state $\varphi$; updating by L\"uders conditionalizing on this
knowledge brings her credences into line with the objective chances
assigned by $\varphi$.
\eje

Because there is no objective connection between raw experiences and
POVM elements, there is no objective post-update state. The remainder
of Earman's treatment of this topic ignores this basic feature of
QBism and is thus inconsequential.

\section{Personalist Probability}
Some of Earman's statements about the conceptual basics of personalist
probability theory are erroneous. For example, Earman asks how
``QBians'' can appeal to a Dutch-book construction \bje when $A$ and
$B$ are noncommuting and there is no possibility of settling
simultaneous bets on $A$, $B$, and $A + B$?  \eje A cursory reading of
the technical papers from the QBist research program would indicate
that Dutch-book arguments are never made across multiple mutually
exclusive experiments. Indeed, recognizing what must be \emph{added}
to Dutch-book arguments to bridge expectations between such
alternatives is \emph{the} central focus of that
work~\cite{Fuchs:2016, Appleby:2017, Stacey:2019}.

\bje
In QBism where the quantum state is merely a device for representing
the credence function of a Bayesian agent, the state changes only when
the agents's {\rm [sic]} credence function changes. Changes in the credence
function can happen because the agent updates her credence function on
new information, or because of some sort of drift in beliefs
uninformed by new information.  QBians do not discuss the latter
possibility, and for good reason since uninformed drift does not have
a rational explanation. The upshot for QBians is that, as far as
rational agents are concerned, there is no Schr\"odinger state evolution
between updating events.
\eje

This is incorrect, because it is founded on a basic misconception
about the treatment of time in personalist probability
theory~\cite{Barzegar:2020}.

To illustrate, think of Alice the weather forecaster.  On Monday,
which we'll call day 0, she makes a forecast for the coming week.  She
writes $P_0(\hbox{rain on day 1})$, $P_0(\hbox{rain on day 2})$ and so
forth.  The subscripts refer to the time at which she sets her
probabilities --- the time at which she makes her gambling
commitments.  The relation between $P_0(\hbox{rain on day 1})$ and
$P_0(\hbox{rain on day 2})$ could be quite complicated, relying upon a
representation of the atmosphere in terms of a stochastic process.
This is not a \emph{change} of probabilities, but rather a listing of
probabilities for multiple related but distinct experiments: testing
for rain on day 1, testing for rain on day 2, etc. \emph{Updating}
probabilities, whether by conditionalization or a more general
probability kinematics, only enters the picture when day 1 rolls
around and Alice must decide how to set her gambling commitments with
a new subscript, like $P_1(\hbox{rain on day 2})$.

The situation in quantum mechanics is exactly analogous. A density
matrix $\rho_0(t = 1)$ encodes Alice's expectations, asserted at time
0, about the potential results of experiments that she could conduct
at time 1. Alice can use the Schr\"odinger equation to calculate
$\rho_0(t = 2)$, by conjugating with a unitary:
\begin{equation}
  \rho_0(t = 2) = U \rho_0(t = 1) U^\dag.
\end{equation}
This is not a change of probabilities, but rather a listing of
probabilities for multiple related but distinct experiments. The fact
that any density matrix is equivalent to a probability distribution
over the outcomes of a reference measurement~\cite{DeBrota:2020,
  Appleby:2017, Slomczynski:2020} underlines that in this respect, the
quantum and classical stories are congruent. Moreover, fixing a
reference measurement also gives a natural representation of unitaries
as \emph{conditional} probabilities, revealing that Schr\"odinger
evolution is simply a deformed counterpart of classical stochastic
evolution~\cite{Fuchs:2010,Fuchs2017}. This mathematical fact is one reason why
QBism explicitly grants the same conceptual status to quantum states
and to time-evolution operators, treating them even-handedly as
doxastic quantities.

Because Earman has an incorrect view of how personalist probability
includes time evolution, he erroneously infers that QBism must abandon
the Schr\"odinger picture in favor of the Heisenberg:
\bje
In the conventional treatment of quantum evolution, Heisenberg and
Schr\"odinger evolution are flip sides of the same coin since
$\omega_0(A_t) = \omega_t(A_0)$. But in QBism Heisenberg evolution has
primacy since the right hand side of this equality makes no sense for
the QBian unless it is understood as a notational variant of the left
hand side.
\eje
For the reasons explained above, this is inaccurate. A QBist can
equally well use the Heisenberg or the Schr\"odinger picture.
\bje
What is disquieting about the QBian stance here is the dualism it
implies: there is something like a realist/objectivist commitment to
the structure of quantum observables and their temporal evolution but
an instrumentalist/subjectivist attitude towards quantum states.
\eje
As explained above, QBism contains no such dualism. Fuchs rejected
this dualism well before ``Quantum Bayesianism'' had matured into
QBism, writing in 2002 that observables and time-evolution operators
are ``subjective information'' just like quantum states~\cite[\S
  7]{Fuchs:2002}.

\section{Conclusion: A Definite Outcome}
Earman's final section before his concluding remarks is a discussion
of whether QBism can be said to resolve conceptual problems in the
interpretation of quantum mechanics. Regarding the question of why
measurements have definite outcomes, he writes,
\bje
As long as a QBian agent is treated as an abstract, disembodied
probability calculator that is fed information by an oracle, the issue
can be avoided. But it resurfaces for physically embodied observers,
such as ourselves, whose information acquisition has to be treated
quantum mechanically in terms of an interaction with the (measurement
apparatus $+$ object system).
\eje
This is a fundamental misrepresentation of how QBists treat measurement in quantum theory:  It ignores what even the Wikipedians know, ``A measurement apparatus is conceptually an extension of the agent. It should be considered analogous to a sense organ or prosthetic limb---simultaneously a tool and a part of the individual.''  The trinary decomposition {\bf object} $+$ {\bf apparatus} + {\bf agent} simply does not exist in QBism. QBism is all about the agent and her external world---the decomposition is a binary one.

In some other interpretations, chiefly
those of an Everettian bent, a measurement involves an observer
becoming entangled with an apparatus that has become entangled with a system. The stories of how exactly this
leads to everyday experience are elaborate and mutually
contradictory~\cite{Kent:2015}. However, at root, formulating an
entangled state for the observer and the system, or the observer and
the system and an apparatus, requires ascribing a quantum state to the
observer. This contradicts the tenets of QBism, according to which an
agent does not assign a quantum state to herself, and there is no
God's-eye super-observer who assigns a state to all
agents~\cite{DeBrota:2020}. QBism indeed regards agents as embodied; how could a disembodied entity take physical actions and experience
consequences? The argument that because agents are embodied their
interactions with the world must be treated as the generation of
entangled states simply presumes its conclusion.

Perhaps we could say more, but this is enough for now.  The scholarship in Ref.\ \cite{Earman:2019} is abysmal: QBians do not exist.

\acknowledgments
CAF and BCS were supported by the John Templeton
Foundation. The opinions expressed in this publication are those of
the authors and do not necessarily reflect the views of the John
Templeton Foundation.  CAF was further supported in part by the John E. Fetzer
Memorial Trust.


\begin{thebibliography}{999}

\bibitem{deFinetti1990}
B. de Finetti, \booktitle{Theory of Probability}, (Wiley, New York, 1990).

\bibitem{Earman:2019} J.\ Earman,
  ``\hrefdoi{10.1093/monist/onz017}{Quantum Bayesianism Assessed},''
  \booktitle{The Monist} \textbf{102} (2019),
  403--23. \url{http://philsci-archive.pitt.edu/14824/}.

\bibitem{Caves02a}
C.~M. Caves, C.~A. Fuchs and R.~Schack, ``\href{https://doi.org/10.1103/PhysRevA.65.022305}{Quantum Probabilities as Bayesian Probabilities},'' \booktitle{Phys.\ Rev.\ A} {\bf 65}, 022305 (2002).

\bibitem{Fuchs:2010} C.\ A.\ Fuchs, ``QBism, the Perimeter of Quantum
  Bayesianism,'' \arxiv{1003.5209} (2010).

\bibitem{FMS}
C.~A. Fuchs, N. D. Mermin, and R.~Schack, ``\href{https://doi.org/10.1119/1.4874855}{An Introduction to QBism with an Application to the Locality of Quantum Mechanics},'' \booktitle{Am.\ J. Phys.}\ {\bf 82}, 749--754 (2014).

\bibitem{FuchsStacey2018}
C. A. Fuchs and B. C. Stacey, ``\href{https://arxiv.org/abs/1612.07308}{QBism:\ Quantum Theory as a Hero's Handbook},'' in {\sl Proceedings of the International School of Physics ``Enrico Fermi'' Course 197 -- Foundations of Quantum Physics}, edited by E.~M. Rasel, W.~P. Schleich, and S. W\"olk (IOS Press, Amsterdam; Societ\`a Italiana di Fisica, Bologna, 2018), pp.\ 133--202.

\bibitem{Stacey:2019b} B.\ C.\ Stacey, ``Ideas Abandoned en Route to QBism,'' \arxiv{1911.07386} (2019).

\bibitem{Earman18}
J. Earman, ``The Relation between Credence and Chance:\ Lewis' `Principal Principle' Is a Theorem of Quantum Probability Theory,'' \href{http://philsci-archive.pitt.edu/14822/}{http://philsci-archive.pitt.edu/14822/} (2018).

\bibitem{Fuchs09}
C.~A. Fuchs and R.~Schack, ``Quantum-Bayesian Coherence,'' \href{https://arxiv.org/abs/0906.2187}{\tt arXiv:0906.2187} (2009).

\bibitem{Fuchs13a}
C.~A. Fuchs and R.~Schack, ``Quantum-Bayesian Coherence,'' \booktitle{Rev.\ Mod.\ Phys.}\ {\bf 85}, 1693--1715 (2013).

\bibitem{Wiki}
``Quantum Bayesianism,'' \href{https://en.wikipedia.org/wiki/Quantum_Bayesianism}{https://en.wikipedia.org/wiki/Quantum$\underline{\phantom{x}}$Bayesianism}, accessed 27 December 2020.

%%%%%%%%%%%%Earman List%%%%%%%%%%%%%%%%%%%%

\bibitem{Caves02c}
C.~M. Caves, C.~A. Fuchs, and R.~Schack, ``Conditions for Compatibility of Quantum-State Assignments,'' \booktitle{Phys.\ Rev.\ A} {\bf 66}, 062111 (2002).

\bibitem{Caves02b}
C.~M. Caves, C.~A. Fuchs and R.~Schack, ``Unknown Quantum States:\ The Quantum de Finetti Representation,'' \booktitle{J. Math.\ Phys.}\ {\bf 43}, 4537--4559 (2002).

\bibitem{Caves07}
C.~M. Caves, C.~A. Fuchs, and R.~Schack, ``Subjective Probability and Quantum Certainty,'' \booktitle{Stud.\ Hist.\ Phil.\ Mod.\ Phys.}\ {\bf 38}, 255--274 (2007).

\bibitem{Fuchs:2002} C.\ A.\ Fuchs, ``Quantum mechanics as quantum
  information (and only a little more),'' \arxiv{quant-ph/0205039}
  (2002).

\bibitem{Fuchs04b}
C.~A. Fuchs and R.~Schack, ``Unknown Quantum States and Operations, a
Bayesian View,'' in \booktitle{Quantum Estimation Theory}, edited by M.~G.~A.
Paris and J. \v{R}eh\'a\v{c}ek, (Springer-Verlag, Berlin, 2004),
pp.~151--190; \arxiv{quant-ph/0404156}.

\bibitem{Fuchs04a}
C.~A. Fuchs, R.~Schack, and P.~F. Scudo, ``A de Finetti
Representation Theorem for Quantum Process Tomography,'' \booktitle{Phys.\
Rev.\ A} {\bf 69}, 062305 (2004).

\bibitem{Mermin12a}
N. D. Mermin, ``Quantum Mechanics:\ Fixing the Shifty Split,'' \booktitle{Phys.\ Today} {\bf 65}(7), 8--10 (2012); reprinted in N.~D. Mermin, \booktitle{Why Quark Rhymes with Pork and other Scientific Diversions}, (Cambridge University Press, Cambridge, UK, 2016), Chapter 31.

\bibitem{Mermin14c}
N. D. Mermin, ``QBism Puts the Scientist Back into Science,'' \booktitle{Nature} {\bf 507}, 421--423 (2014).

\bibitem{Schack01a}
R. Schack, T. A. Brun, and C. M. Caves, ``Quantum Bayes Rule,'' \booktitle{Phys.\ Rev.\ A} {\bf 64}, 014305 (2001).

\bibitem{vonBaeyer16a}
H. C. von Baeyer, {\sl QBism:\ The Future of Quantum Physics}, (Harvard University Press, Cambridge, MA, 2016).

\bibitem{Timpson08a}
C. G. Timpson, ``Quantum Bayesianism:\ A Study,'' \booktitle{Stud.\ Hist.\ Phil.\ Mod.\ Phys.}\ {\bf 39}, 579--609 (2008).

%%%%%%%%%%%%%%%%%%%%%%%%%%%%%%%%%%%%%%%%%%%

\bibitem{Samizdat2}
C.~A. Fuchs, {\sl My Struggles with the Block Universe:\ Selected Correspondence, January 2001 -- May 2011}, edited by Blake C. Stacey, foreword by Maximilian Schlosshauer (2014), 2,349 pages; \href{https://arxiv.org/abs/1405.2390}{\tt arXiv:1405.2390}.

\bibitem{Fuchs20}
C. A. Fuchs and B. C. Stacey, ``Are Non-Boolean Event Structures the Precedence or Consequence of Quantum Probability?,'' to appear in \booktitle{Probing the Meaning and Structure of Quantum Mechanics}, edited by D. Aerts, J.~B. Arenhart, C. de~Ronde, and G. Sergioli (World Scientific, Singapore, 2020); \href{https://arxiv.org/abs/1912.10880}{\tt arXiv:1912.10880}.

\bibitem{Fuchs11}
C.~A. Fuchs and R.~Schack, ``\href{https://doi.org/10.1007/s10701-009-9404-8}{A Quantum-Bayesian Route to Quantum-State Space},'' \booktitle{Found.\ Phys.}\ {\bf 41}, 345--356 (2011).

\bibitem{Appleby11b}
D.~M. Appleby, {\AA}.~Ericsson, and C.~A. Fuchs, ``\href{https://doi.org/10.1007/s10701-010-9458-7}{Properties of QBist State Spaces},'' \booktitle{Found.\ Phys.}\ {\bf 41}, 564--579 (2011).

\bibitem{Appleby11c}
D. M. Appleby, S. T. Flammia, and C. A. Fuchs, ``\href{https://doi.org/10.1063/1.3555805}{The Lie Algebraic Significance of Symmetric Informationally Complete Measurements},'' \booktitle{J. Math.\ Phys.}\ {\bf 52}, 022202 (2011).

\bibitem{Appleby13}
D.~M. Appleby, H.~B. Dang, and C.~A. Fuchs, ``\href{https://doi.org/10.3390/e16031484}{Symmetric Informationally-Complete Quantum States as Analogues to Orthonormal Bases and Minimum-Uncertainty States},'' \booktitle{Entropy} {\bf 16}, 1484--1492 (2014).

\bibitem{Appleby15}
D. M. Appleby, C. A. Fuchs, and H. Zhu, ``Group Theoretic, Lie Algebraic and Jordan Algebraic For\-mu\-lations of the SIC Existence Problem,'' \booktitle{Quant.\ Info.\ Comp.}\ {\bf 15}, 61--94 (2015), \arxiv{1312.0555}.

\bibitem{Appleby:2017} M.\ Appleby, C.\ A.\ Fuchs, B.\ C.\ Stacey and
  H.\ Zhu, ``\hrefdoi{10.1140/epjd/e2017-80024-y}{Introducing the
    Qplex:\ A novel arena for quantum theory},'' \booktitle{European
    Physical Journal D} \textbf{71} (2017), 197, \arxiv{1612.03234}.
    
\bibitem{DeBrota20b}
J. B. DeBrota, C. A. Fuchs, and B. C. Stacey, ``\href{https://doi.org/10.1103/PhysRevResearch.2.013074}{Symmetric Informationally Complete Measurements Identify the Irreducible Difference between Classical and Quantum Systems},'' Phys.\ Rev.\ Research {\bf 2}, 013074 (2020).
    
\bibitem{Fuchs11b}
C. A. Fuchs, ``Interview with a Quantum Bayesian,'' in {\sl Elegance and Enigma:\ The Quantum Interviews}, edited by M.~Schlosshauer (Springer, Berlin, Frontiers Collection, 2011), \arxiv{1207.2141}.

\bibitem{Fuchs14}
C. A. Fuchs and R. Schack, ``Quantum Measurement and the Paulian Idea,'' in \booktitle{The Pauli-Jung Conjecture and Its Impact Today}, edited by H. Atmanspacher and C. A. Fuchs (Imprint Academic, Exeter, UK, 2014), pp.\ 93--107, \arxiv{1412.4209}.

\bibitem{DeBrota:2020}
J. B. DeBrota, C. A. Fuchs, and R. Schack, ``\href{https://doi.org/10.1007/s10701-020-00369-x}{Respecting One's Fellow:\ QBism's Analysis of Wigner's Friend},'' \booktitle{Found.\ Phys.}\ {\bf 50}, 1859--1874 (2020).

\bibitem{Fuchs15b}
C. A. Fuchs and R. Schack, ``\href{https://iopscience.iop.org/article/10.1088/0031-8949/90/1/015104}{QBism and the Greeks:\ Why a Quantum State Does Not Represent an Element of Physical Reality},'' \booktitle{Physica Scripta} {\bf 90}, 015104 (2015).

\bibitem{Fuchs2017}
C. A. Fuchs, ``Notwithstanding Bohr, the Reasons for QBism,'' \booktitle{Mind and Matter} {\bf 15}, 245--300 (2017); \href{https://arxiv.org/abs/1705.03483}{\tt arXiv:1705.03483}.

\bibitem{Wheeler:1982} J.\ A.\ Wheeler, ``Bohr, Einstein, and the
  strange lesson of the quantum.'' In \booktitle{Mind in
    Nature:\ Nobel Conference XVII, Gustavus Adolphus College,} edited
  by R.\ Q.\ Elvee. (Harper \& Row, 1982.)
  
\bibitem{Fuchs16}
C. A. Fuchs, ``On Participatory Realism,'' in \booktitle{Information and Interaction:\ Eddington, Wheeler, and the Limits of Knowledge}, edited by I.~T. Durham and D. Rickles, (Springer, Berlin, 2016), pp.\ 113--134, \arxiv{1601.04360}.

\bibitem{Whitehead:1979} A.\ N.\ Whitehead, \booktitle{Process and
  Reality:\ An Essay in Cosmology,} edited by D.\ R.\ Griffin and
  D.\ W.\ Sherburne. (Free Press, 1979.)

\bibitem{Barad:2007} K.\ Barad, \booktitle{Meeting the Universe
  Halfway:\ Quantum Physics and the Entanglement of Matter and
  Meaning}. (Duke University Press, Durham, NC, 2007.)

\bibitem{Busch:2003} P.\ Busch,
  ``\hrefdoi{10.1103/PhysRevLett.91.120403}{Quantum States and
  Generalized Observables:\ A Simple Proof of Gleason's Theorem},''
  \booktitle{Physical Review Letters} \textbf{91} (2003), 120403,
  \arxiv{quant-ph/9909073}.

\bibitem{Caves:2004} C.\ M.\ Caves, C.\ A.\ Fuchs, K.\ K.\ Manne and
  J.\ M. Renes,
  ``\hrefdoi{10.1023/B:FOOP.0000019581.00318.a5}{Gleason-Type
    Derivations of the Quantum Probability Rule for Generalized
    Measurements},'' \booktitle{Foundations of Physics} \textbf{34}
  (2004), 193--209, \arxiv{quant-ph/0306179}.

\bibitem{Barnum:2000} H.\ Barnum, ``Information-disturbance tradeoff
  in quantum measurement on the uniform ensemble and on the mutually
  unbiased bases,'' \arxiv{quant-ph/0205155} (2000).

\bibitem{Busch:2009} P.\ Busch and P.\ Lahti,
  ``\hrefdoi{10.1007/978-3-540-70626-7_110}{L\"uders Rule}.'' In
  \booktitle{Compendium of Quantum Physics} (Springer,
  2009). \url{http://philsci-archive.pitt.edu/4111/}.
  
\bibitem{Hacking67}
I. Hacking, ``Slightly More Realistic Personal Probability,'' \booktitle{Phil.\ Sci.}\ {\bf 34}, 311 (1967).

\bibitem{vanFraassen84}
B. C. van Fraassen, ``Belief and the Will,'' \booktitle{J. Phil.}\ {\bf 81}, 235 (1984).

\bibitem{Fuchs:2012} C.\ A.\ Fuchs and R.\ Schack,
  ``\hrefdoi{10.1007/978-3-642-21329-8_15}{Bayesian Conditioning, the
  Reflection Principle, and Quantum Decoherence}.'' In
  \booktitle{Probability in Physics,} edited by Y.\ Ben-Menahem and
  M.\ Hemmo. (Springer, 2012.) \arxiv{1103.5950}.

\bibitem{DeBrota:2019} J.\ B.\ DeBrota and B.\ C.\ Stacey,
  ``\hrefdoi{10.1103/PhysRevA.100.062327}{L\"uders Channels and the
  Existence of Symmetric Informationally Complete Measurements},''
  \booktitle{Physical Review A} \textbf{100} (2019), 062327,
  \arxiv{1907.10999}.

\bibitem{Wilde:2017} M.\ M.\ Wilde, \booktitle{Quantum Information
  Theory}, second edition (Cambridge University Press,
  2017). \arxiv{1106.1445}.

\bibitem{Ruskai:2003} M.\ B.\ Ruskai,
  ``\hrefdoi{10.1142/S0129055X03001710}{Qubit Entanglement Breaking
  Channels},'' \booktitle{Reviews in Mathematical Physics} \textbf{15}
  (2003), 643--62, \arxiv{quant-ph/0302032}.

\bibitem{DeBrota:2020b} J.\ B.\ DeBrota, C.\ A.\ Fuchs and
  B.\ C.\ Stacey,
  ``\hrefdoi{10.1103/PhysRevResearch.2.013074}{Symmetric
    informationally complete measurements identify the irreducible
    difference between classical and quantum systems},''
  \booktitle{Physical Review Research} \textbf{2} (2020), 013074,
  \arxiv{1805.08721}.

\bibitem{Pandey:2020} S.\ K.\ Pandey, V.\ I.\ Paulsen, J.\ Prakash and
  M.\ Rahaman, ``\hrefdoi{10.1063/1.5045184}{Entanglement Breaking
    Rank and the Existence of SIC POVMs},'' \booktitle{Journal of
    Mathematical Physics} \textbf{61} (2020), 042203,
  \arxiv{1805.04583}.

\bibitem{Fuchs:2016} C.\ A.\ Fuchs and B.\ C.\ Stacey, ``Some Negative
  Remarks on Operational Approaches to Quantum Theory.''  In
  \booktitle{Quantum Theory:\ Informational Foundations and Foils,}
  edited by G.\ Chiribella and R.\ W.\ Spekkens. (Springer, 2016.)
  \arxiv{1401.7254}.

\bibitem{Stacey:2019} B.\ C.\ Stacey, ``Quantum Theory as Symmetry
  Broken by Vitality,'' \arxiv[quant-ph]{1907.02432} (2019).

\bibitem{Barzegar:2020} A.\ Barzegar,
  ``\hrefdoi{10.1007/s10701-020-00347-3}{QBism is Not So Easily
  Dismissed},'' \booktitle{Foundations of Physics} \textbf{50} (2020),
  693--707, \arxiv{2006.02790}.

\bibitem{Slomczynski:2020} W.\ S{\l}omczy\'nski and A.\ Szymusiak,
  ``\hrefdoi{10.22331/q-2020-09-30-338}{Morphophoric POVMs,
  generalised qplexes, and 2-designs},'' \booktitle{Quantum}
  \textbf{4} (2020), 338, \arxiv{1911.12456}.

\bibitem{Kent:2015} A.\ Kent,
  ``\hrefdoi{10.1007/s10701-014-9862-5}{Does it make sense to speak of
  self-locating uncertainty in the universal wave function?  Remarks
  on Sebens and Carroll},'' \booktitle{Foundations of Physics}
  \textbf{45} (2015), 211--17, \arxiv{1408.1944}.

\end{thebibliography}
\end{document}